\documentclass[10pt, journal]{IEEEtran}
\usepackage[dvips]{graphicx}
\usepackage{amsmath}
\usepackage{subfigure}
\usepackage{epsfig, psfrag, amsmath, amssymb, amsfonts, latexsym, balance,indentfirst,}
\usepackage[numbers,sort&compress]{natbib}
\usepackage{algorithm, algorithmic}
\usepackage{multirow}
\usepackage{amsthm}
\usepackage{xcolor}
\usepackage{extarrows}
\usepackage[small]{caption2}

\usepackage{booktabs}
\usepackage{makecell}

\allowdisplaybreaks
\newlength{\figwidth}
\ifCLASSOPTIONonecolumn
\usepackage{geometry}
\usepackage{lipsum}
\else
\fi
\begin{document}

\title{Real Entropy Can Also Predict Daily Voice Traffic for Wireless Network Users}

\author{
	\IEEEauthorblockN{
		Sihai Zhang\IEEEauthorrefmark{1},
        Junyao Guo\IEEEauthorrefmark{1},
		Tian Lan\IEEEauthorrefmark{1},
		Rui Sun\IEEEauthorrefmark{1},
		Jinkang Zhu\IEEEauthorrefmark{2}}

	\IEEEauthorrefmark{1}Key Laboratory of Wireless-Optical Communications, \\
	University of Science \& Technology of China, Hefei, Anhui, P.R. China\\
	\IEEEauthorrefmark{2} PCNSS, University of Science \& Technology of China, Hefei, Anhui, P.R. China\\
	
	Email: shzhang@ustc.edu.cn, \{gjyh, lantian1, issunrui\}@mail.ustc.edu.cn, jkzhu@ustc.edu.cn	
}

	
	

%
\maketitle
\thispagestyle{empty}
\begin{abstract}
	
Voice traffic prediction is significant for network deployment optimization thus to improve the network efficiency. The real entropy based theorectical bound and corresponding prediction models have demonstrated their success in mobility prediction. In this paper, the real entropy based predictability analysis and prediction models are introduced into voice traffic prediction. For this adoption, the traffic quantification methods is proposed and discussed. Based on the real world voice traffic data, the prediction accuracy of N-order Markov models, diffusion based model and MF model are presented, among which, 25-order Markov models performs best and approach close to the maximum predictability. This work demonstrates that, the real entropy can also predict voice traffic well which broaden the understanding on the real entropy based prediction theory.

\end{abstract}


\section{Introduction}

{
Traffic modelling and prediction has been an interesting difficult issue in mobile communication networks.
Facing the continuous increasing of subscribers and traffic load, operators nowadays have to satisfy the QoS (Quality of Service) demand of subscribers\cite{consumption}.
So in the past few years, cellular traffic patterns have been extensively studied.
Lee \cite{Spatialmodeling} revealed that traffic density in spatial domain can be best approximated by Log-normal distribution or Weibull distribution.
Wang et al.\cite{Characterizing} found that mobile traffic follows a trimodal distribution, which is the combination of compound-exponential, power-law and exponential distributions, in both spatial and temporal dimensions.
In addition, Shafiq et al.\cite{Characterizinggeospatial} characterized and simulated the Internet traffic dynamics of cell traces.
\cite{Bigdata} studied the regularity and randomness of traffic patterns in millions of cellular towers by time series analysis.
However, current understanding about the mobile traffic patterns is still limited, which significantly increases the cost of operating millions of cellular towers in big cities and reduces the quality of service provided.
}

{
As to traffic prediction, there are user traffic prediction, base station traffic prediction\cite{Erlang} and large call center traffic prediction.
Several common traffic prediction methods include ARIMA (Autoregressive Integrated Moving Average model), SVM (support vector machine), neural network\cite{Forecasting}, markov prediction model, naive bayes classifier and logistic regression\cite{usertraffic}.
LSTM (Long Short-Term Memory) model is also used to study the mobile traffic of LTE base station, and the one-step prediction and long-term prediction errors are evaluated\cite{MobileTrafficLSTM}.

{
In response to the deficiencies of existing research, this paper introduces the real entropy concepts which originates in mobility prediction into the traffic prediction, and expects the prediction accuracy of voice traffic can be explained with stronger theorectical supports.
We also propose a traffic quantification method to balance the complexity of state space with the prediction accuracy by choosing different values for quantification parameter.
So our work process the detailed call records of a certain city provided by the operator and predict the next day voice traffic of wireless users.
In addition, this paper discusses the tradeoff between the complexity of traffic model and the prediction accuracy by combining theoretical analysis with practical prediction algorithm.
}

The contributions of this paper are as following:

\begin{itemize}
{
\item We introduce the real entropy concept which originates in the discrete mobility prediction, to characterize the uncertainty of the user's voice traffic.
Using the quantification methods with different interval parameters, the transition state space of voice traffic are estabilished and the entropy together with predictability can be obtained, which present the theorectical guide for voice traffic prediction.
\item N-order Markov transition models, diffusion based model and MF model are used to predict the voice traffic of nearly 200,000 users in the real data set.
 We compare the predicted results with the theoretical values.
 The results show that user traffic is highly dependent on historical states, and the maximum predictability is not only the underlying theoretical limit, but also the accessible target of the actual algorithm prediction accuracy.
 Under the coarse-grained quantization of T=600s, the prediction accuracy can be as high as 83.47\%.
  }

\end{itemize}

\section{Concepts and Model}
\label{sec:con}

\subsection{Definition of Real Entropy}
Generally speaking, the prediction technique is utilized to foresee the state of target object at some point in the future according to its historical time series.
The time series consists of state information in chronological order, which can be denoted as $X={s_1,s_2,...,s_n}$. Entropy is a useful quantity measuring the degree of predictability of time series. Considering the different level of state information, it can derive three kinds of entropy to describe the fundamental predictability of the state corresponding to the target object \cite{predictability}.

\emph{i) Random entropy}: $S^{\text{rand}}=\rm{log}_{2}N$, where $N$ is the number of unique states that the target object had ever been.

\emph{ii) Temporal-uncorrelated entropy}:  $S^{\text{unc}}=-\sum_{i}p(s_{i})\rm{log}_2p(s_{i})$, where $p(s_{i})$ is the probability that state $s_{i}$ occurs in the history.

\emph{iii) Real entropy}: Capturing the temporal relationship among the different states, the real entropy can be deduced. To be specific, for a target object with $X={s_1,s_2,...,s_n}$, the real entropy $S^{\text{real}}$ can be given by $-\sum_{X^{\prime}\subset
	X}p(X^{\prime}_{i})\rm{log} _{2}p(X^{\prime}_{i})$, where $p(X^{\prime}_{i})$ is the probability of finding a particular time-ordered subsequence $X^{\prime}$.

\subsection{Definition of Predictability}
The probability $\Pi$ that an appropriate predictive algorithm can predict correctly the user¡¯s future whereabouts is an important measure of predictability. It can be related to entropy calculated based on records of object state using a version of Fano's equality \cite{predictability}.

Given the entropy $S$ for an object who has ever been $N$ unique states, its predictability probability $\Pi\leq\Pi^{\text{max}}$, where $\Pi^{\text{max}}$ can be given by $S = H(\Pi^{\text{max}})+(1-\Pi^{\text{max}})\rm{log}_2(N-1)$ and $H(\Pi^{\text{max}}) = -\Pi^{\text{max}}log_2(\Pi^{\text{max}})-(1-\Pi^{\text{max}})\rm{log}_2(\Pi^{\text{max}})$. For a target object with $\Pi^{max}=0.5$, it means that only in the 50\% of time the state of the target object can be hoped to foresee correctly.

Let $\Pi^{\text{rand}} = \Pi^{\text{rand}}(S^{\text{rand}},N)$, $\Pi^{\text{unc}} = \Pi^{\text{unc}}(S^{\text{unc}},N)$ and $\Pi^{\text{real}} = \Pi^{\text{real}}(S^{\text{real}},N)$. Since $S^{\text{rand}}\geq S^{\text{unc}}\geq S^{\text{real}}$, it is true that $\Pi^{\text{rand}}\geq \Pi^{\text{unc}}\geq \Pi^{\text{real}}$. Through this quantity, it is effective to measure the inherent predictive power of the records.

\subsection{Definition of States in Voice Traffic Prediction}
\label{subsec:quantification}

{When the mobility prediction is concerned, the states of users are naturally and easily decided, which is the geographic locations, like the areas coverd by base stations, the buildings visited by users and etc.
When comes to the daily voice traffic prediction, the proper state space should be considerd.
In this paper, the parameter of traffic interval $T$ is adopted to process the CDR data and generate the voice traffic time series for predictions.

As depicted in Fig. \ref{fig-mapping}, the meaning of $T$ is the quantization interval of daily voice traffic.
Since the voice traffic in call detail records (CDR) is measured in seconds, quantization is used to discretize continuously varying traffic, and to approximate the continuous value of the traffic (or a large number of possible discrete values) to a finite number (or fewer) of discrete states.
And each state represents an interval of traffic volume.
Larger $T$ leads to larger quantized granularity with smaller number of states, which may leads to higher accuracy of the predicted traffic state, but the larger the error range compared to the actual voice traffic flow.}

\begin{figure}
	\centerline{\includegraphics[width=0.5\textwidth]{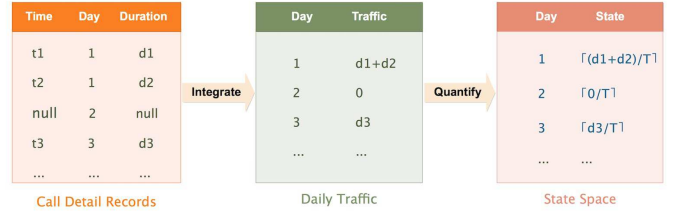}}
	\caption{Processing calling records into daily voice traffic and quantification into states space for one user.}
	\label{fig-mapping}
\end{figure}


\subsection{N-order Markov Prediction Model}

In the existing literatures, $n$-order Markov chain model is an effective prediction approach to perform next state prediction, which predicts the next state using the historical $n-1$ states. The transition probability is calculated by the Eq.\ref{eq4}.

\begin{equation}
\begin{aligned}
\label{eq4}
&P(X^{t+1}=s_{t+1}|X^{t}=s_{t},...,X^{1}=s_{1})\quad\quad\quad\quad\\
&=P(X^{t+1}=s_{t+1}|X^{t}=s_{t},X^{t-n+1}=s_{t-n+1}).
\end{aligned}
\end{equation}

where $X^{t}$ represents the location of one user at time $t$.

According to the transition matrix, the location with the maximum transition probability is chosen as the predicted next location. The prediction accuracy to assess the performance is defined as the number of correct predictions $n_{\text{correct}}$ over the total number of predictions $n_{\text{total}}$:

\begin{equation}\label{eq10}
Accuracy = \frac{n_{\text{correct}}}{n_{\text{total}}}
\end{equation}

\section{Data Sets}
\label{sec:data}

\subsection{Basic Information}

The user traffic data analyzed and predicted in this paper is provided by Hefei Telecom operators and converted from the CDR data set.
The CDR dataset was measured for a total of 184 days from July 1, 2014 to December 31, 2014. The data set contains all the call records of 194,336 users in Hefei during the observation period. The number of records is 247,282,244, and the record format is shown as below:
(SERVICE\underline{ }NBR, CALL\underline{ }TYPE, OPPOSITE\underline{ }NO, TOLLTYPE\underline{ }ID, ROAM\underline{ }TYPE, START\underline{ }TIME, END\underline{ }TIME, DURATION, CITY\underline{ }ID, ROAM\underline{ }CITY\underline{ }ID, OPPCITY\underline{ }ID, LAC\underline{ }ID and CELL\underline{ }ID)

In order to extract the traffic data of users, we take the five features as following: SERVICE\underline{ }NBR, OPPOSITE\underline{ }NO, START\underline{ }TIME, END\underline{ }TIME, DURATION for predictability calculation and next state prediction.

SERVICE\underline{ }NBR and OPPOSITE\underline{ }NO represent the local number and the opposite end number respectively.
 To ensure the privacy of the user, the related number has been anonymized and guaranteed to be unique. START\underline{ }TIME and END\underline{ }TIME indicate the start time and deadline of the user's call, respectively, with a total of 14 digits and accurate to the second. For example, "20141017154209" means 15:42:09 on October 17, 2014. The last row DURATION indicates the duration of the call whose unit is seconds.

\subsection{Data Processing}
{
In order to analyze and predict the user's voice traffic, this paper calculates the daily call duration of each user based on the CDR data set, a total of 180 days of data.
Since the data set contains call records of different users at different timestamp, this paper first merges the call durations of the same date of each user.
It is worth noting that if a user does not have call data for one day, the duration of the call for that day is 0, and a zero value is inserted.
{
The daily voice traffic of about 200,000 users in 184 days is counted, and the maximum value is 80,204 seconds. The daily traffic volume of 90\% users is distributed between 0 $\sim$ 3,780 seconds.
Since the user's per call is measured in seconds, which is approximately a continuous time series, we quantified the traffic volume and mapped it to the designed state space to calculate entropy and predictability.
Considering the user's daily traffic distribution range and the number of traffic states, this paper chooses three quantitative intervals for discussion.
The length of quantization interval $T$ is 120s, 300s and 600s, i.e. 2 minutes, 5 minutes, 10 minutes respectively.}

Fig. \ref{oneuserstate} shows the daily voice traffic of two typical users.
We use the ADF test(Augmented Dickey-Fuller test) method to analyze the stability of the traffic sequence.
{When the $p$ value, the probability corresponding to the $t$-statistic, is less than a given significance level, generally 0.05, 0.01, etc., it means that the sequence is stationary.}
The $p$ value of the two users in Fig. \ref{oneuserstate} is far less than 0.01, so the users' traffic sequence can be seen as stable. In addition, we analyze the users of the entire data set and find that 91.9\% of the user traffic sequences are stable.

\begin{figure}
	\centering
	\subfigure[]{
		\begin{minipage}[b]{0.45\textwidth}
			\includegraphics[width=1\textwidth]{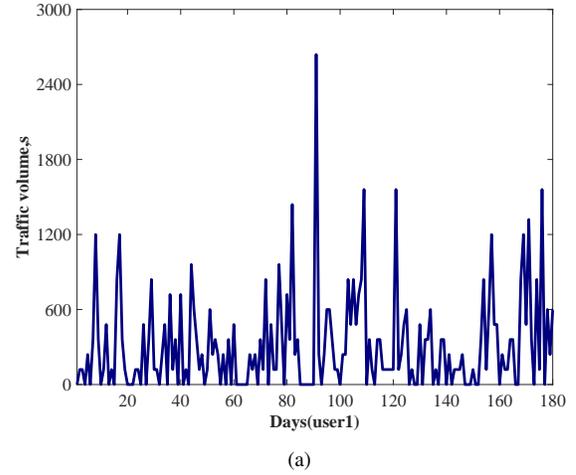}
		\end{minipage}
	}
	\subfigure[]{
		\begin{minipage}[b]{0.45\textwidth}
			\includegraphics[width=1\textwidth]{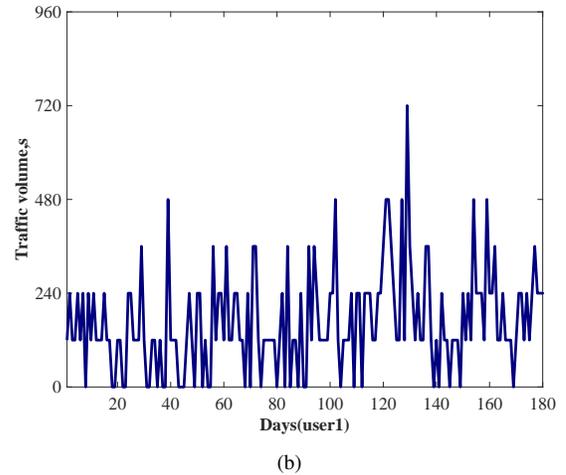}
		\end{minipage}
	}
	\caption{(a) The duration of the call has no obvious periodicity, and the p value of the ADF test is $1.07\times 10^{-23}$. (b) The duration of the call has some periodicity, and the p value of the ADF test is $5.12\times 10^{-11}$. }
	\label{oneuserstate}
\end{figure}


}

{
Fig. \ref{dis_states} shows the distribution of the amount of each users' valid traffic states $N$ for different quantization intervals.
Here, the effective state means that the frequency of users' occurrence of this state is not zero.
As can be seen from Fig. \ref{dis_states} that as the quantization interval increases, the number of effective traffic states of most users decreases, which may lead to better predictive performance.
}

\begin{figure}
\centerline{\includegraphics[width=0.45\textwidth]{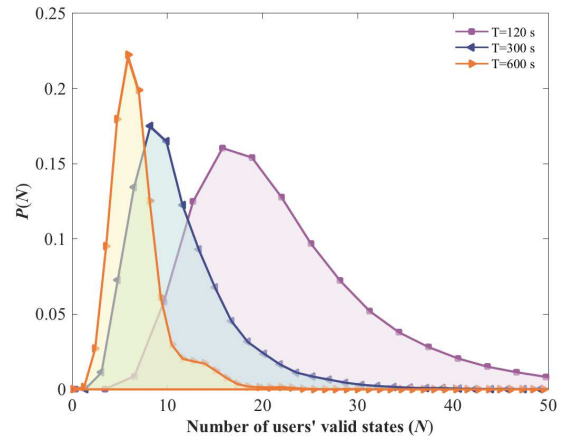}}
\caption{The distribution of number of users' valid traffic states under of different $T$.}
\label{dis_states}
\end{figure}

\section{Actual Entropy and Predictability}

{
Based on the fundamentals of information entropy, the concept $\Pi^{\text{max}}$ given by entropy $S$ characterizes the predictability of spatio-temporal sequence.
The three types of entropy and the corresponding predictability with $T$ = 120 s are shown in Fig. \ref{figpredicta}(a) and Fig. \ref{figpredicta}(b), respectively.

Among them, the difference of $P(S^{\text{rand}})$ and $P(S^{\text{real}})$ is very significant. In fact the peak of random entropy is about $4.3$, which means that each update of user's traffic state represents 5.3 bits new information on average and the next state of the user can be found in $2^{4.3} \approx 20$ random states.
The true entropy peaks at about $2.2$, which means that the user traffic is more predictable after considering the complete spatio-temporal sequence.

The upper limit of predictability $\Pi$ can be obtained according to Fano's inequality \cite{predictability}.
Considering $P(\Pi^{\text{rand}})$, the peak is close to 0, and the prediction using random entropy is basically invalid.
The $P(\Pi^{\text{unc}})$ is widely distributed and reaches a peak at about 0.42.
This indicates that the prediction of user traffic is not ideal by the access frequency only.
 $P(\Pi^{\text{max}})$ reaches a peak at about 0.65 and the range is quite narrow.
 Note that the predictability $\Pi$ of the following text refers to $\Pi^{\text{max}}$.
}

\begin{figure}
	\centering
	\subfigure[]{
		\begin{minipage}[b]{0.45\textwidth}
			\includegraphics[width=1\textwidth]{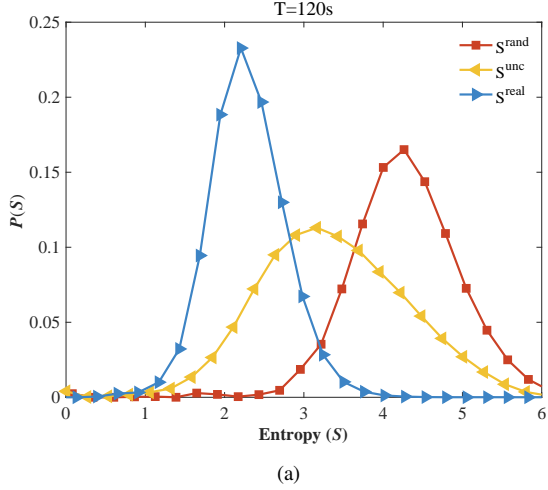}
		\end{minipage}
	}
	\subfigure[]{
		\begin{minipage}[b]{0.45\textwidth}
			\includegraphics[width=1\textwidth]{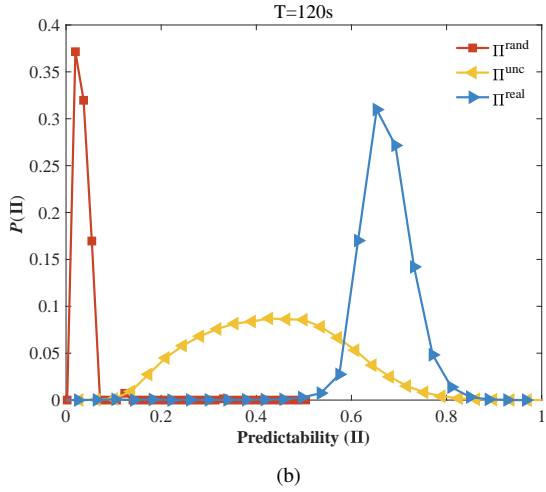}
		\end{minipage}
	}
	\caption{(a) Distribution of three entropies of users' traffic when $T=120s$. (b) Predictability distribution of users' traffic when $T=120s$. }
	\label{figpredicta}
\end{figure}

{
We then calculate and compare the upper bounds of the actual entropy and predictability for the three different quantization intervals, $T$=120s, 300s, and 600s, as shown in Fig. \ref{figpredict_diff}.
It can be seen that the real entropy of users decreases with the increase of quantization interval, and the maximum predictability increases accordingly. When $T$=600s, the maximum predictability reaches about 80.3\%, which implies that the practical prediction accuracy may reach such accuracy.
}
\begin{figure}
	\centering
	\subfigure[]{
		\begin{minipage}[b]{0.45\textwidth}
			\includegraphics[width=1\textwidth]{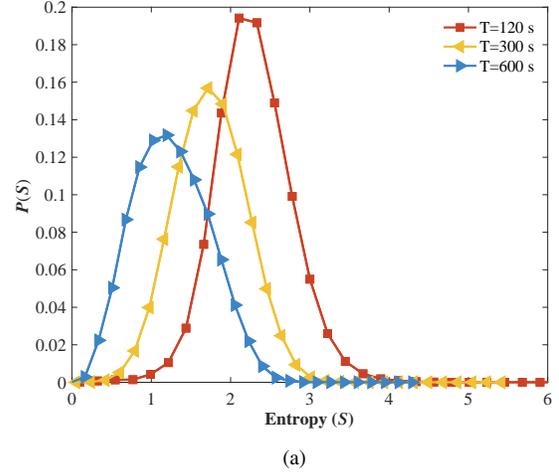}
		\end{minipage}
	}
	\subfigure[]{
		\begin{minipage}[b]{0.45\textwidth}
			\includegraphics[width=1\textwidth]{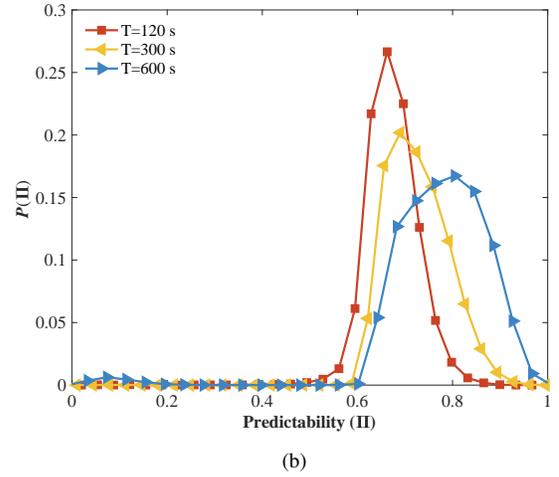}
		\end{minipage}
	}
	\caption{(a) The distribution of real entropy under different time scales. (b)The distribution of maximum predictability under different time scales. }
	\label{figpredict_diff}
\end{figure}

Below, we analyze the theoretical values with appropriate data prediction algorithms.

\section{Prediction Performance and Comparison}
\label{sec4}
Based on the real user traffic data set, this section makes a comparative analysis between the actual prediction accuracy and the theoretical predictability upper bound, which verifies the effectiveness of the predictability measurement in traffic prediction.

\begin{figure}
	\centering
	\subfigure[]{
		\begin{minipage}[b]{0.5\textwidth}
			\includegraphics[width=1\textwidth]{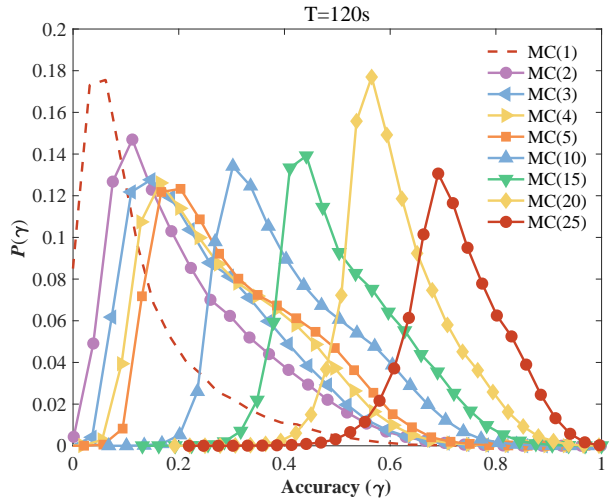}
		\end{minipage}
	}
	\subfigure[]{
		\begin{minipage}[b]{0.5\textwidth}
			\includegraphics[width=1\textwidth]{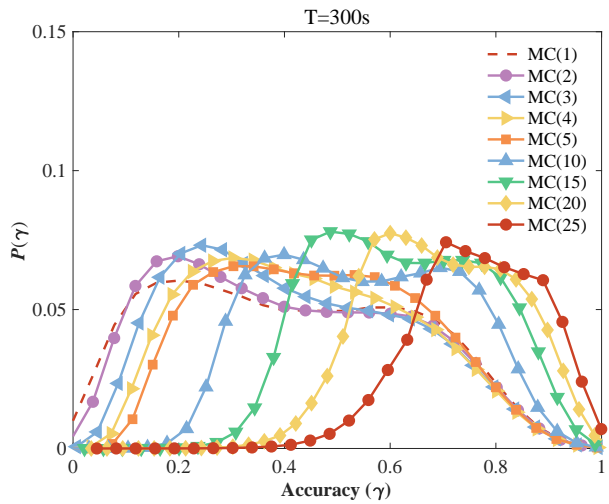}
		\end{minipage}
	}
	\subfigure[]{
		\begin{minipage}[b]{0.5\textwidth}
			\includegraphics[width=1\textwidth]{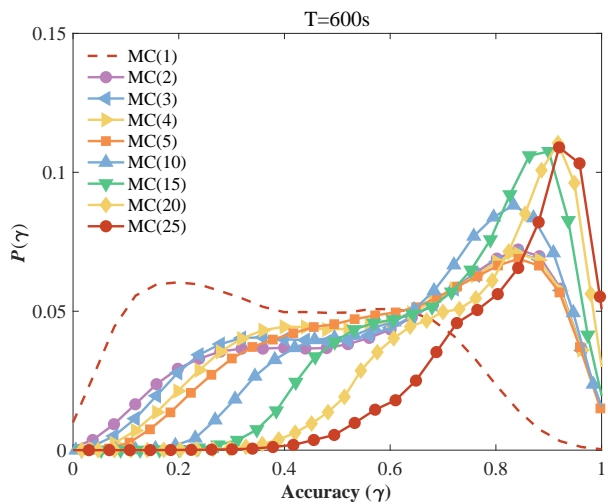}
		\end{minipage}
	}
	\caption{The distribution of prediction accuracy of users' traffic. (a) $T=120s$. (b) $T=300s$. (c) $T=600s$. }
	\label{fig-markov}
\end{figure}

Fig. \ref{fig-markov} presents the probability distribution of prediction accuracy using multi-order Markov model on traffic forecasting.
Firstly, the performance of the low-order Markov model is poor because low-order Markov model cannot recognize the long term correlation in the traffic data.
Secondly, as the order increases, the prediction performance gets better and better but the computational cost of high-order Markov models is also higher. Higher-order models mean that the current state is affected by more previous states, and the correlations used are more prolonged. When the model's order reaches 25, the performance reaches 0.71, which approximates the upper bound of predictability.
Thirdly, from Fig. \ref{fig-markov}(a), (b) to (c), the quantization interval  $T$ increases from 120s, 300s to 600s, and the prediction accuracy also increases, with corresponding peak value gradually shifts to higher accuracy.
\begin{figure}
	\centering
	\subfigure[]{
		\begin{minipage}[b]{0.5\textwidth}
			\includegraphics[width=1\textwidth]{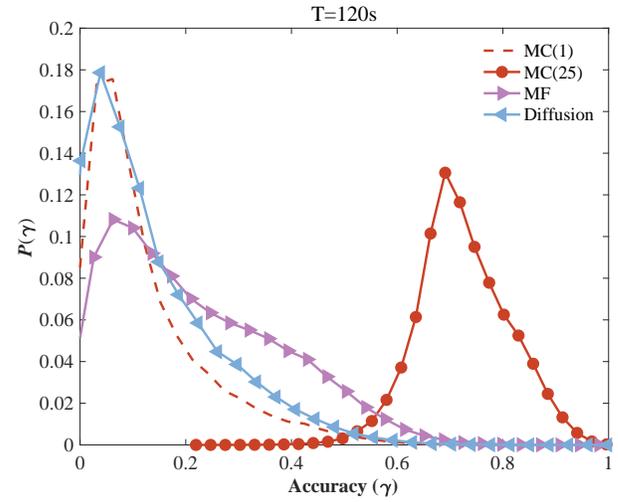}
		\end{minipage}
	}
	\subfigure[]{
		\begin{minipage}[b]{0.5\textwidth}
			\includegraphics[width=1\textwidth]{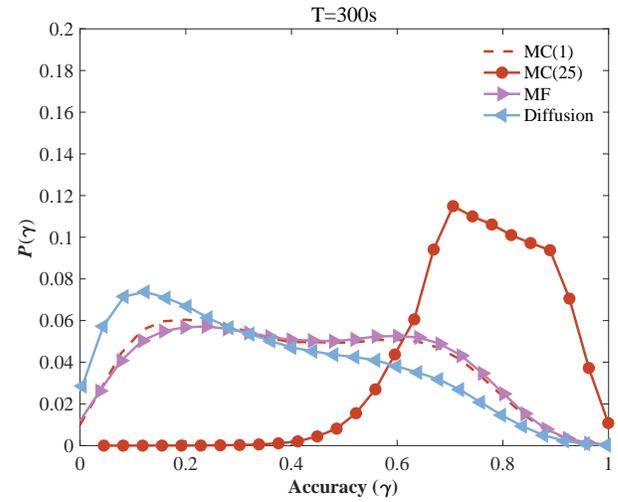}
		\end{minipage}
	}
	\subfigure[]{
		\begin{minipage}[b]{0.5\textwidth}
			\includegraphics[width=1\textwidth]{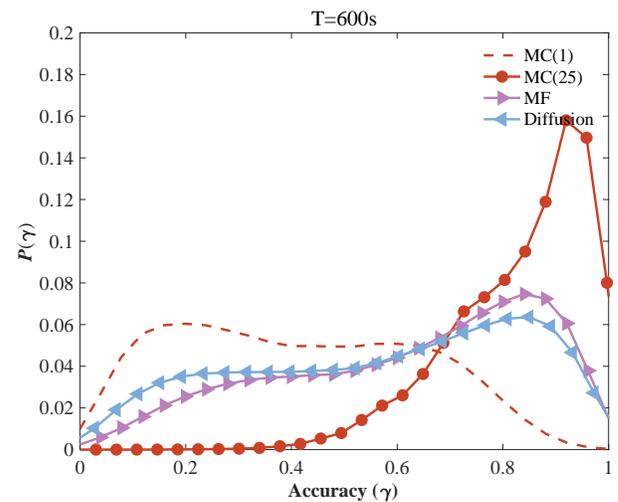}
		\end{minipage}
	}
	\caption{ MC(1) and MC(25) \cite{approachlimit,markovchains} represent the 1-order and 25-order Markov chains. Others are the MF (Most Frequent) model \cite{friendshipandmobility} and the Diffusion Kernel model\cite{diffusion}. The distribution of prediction accuracy of users' traffic. (a) $T=120s$. (b) $T=300s$. (c) $T=600s$. }
	\label{fig-compare}
\end{figure}

In this paper, except for Markov model, other time series prediction models with rosy performance in the field of mobility prediction are also used to predict users' voice traffic. The results of prediction performance comparison are shown in Fig.\ref{fig-compare}.
Intuitively, MC(25) has the best performance and is superior to other predictive models because of exploiting the long correlation and self-similarity of voice traffic.

{
In order to verify the effectiveness of the proposed prediction methods, the performance reported in the current literature are used for comparison.
The ARIMA model \cite{Bigdata} achieves the prediction accuracy = 80\% with the maximum prediction error = 0.3 erlang and the PSO-VSR (Particle swarm optimization-Support vector regression) method\cite{SVR} reaches the prediction accuracy = 80\% with the maximum prediction error = 0.21 erlang.
The performance in this paper is approximately 83.47\% accuracy with an error range of 0.17 Erlang, so the accuracy is excellent when the error range is acceptable, which confirms the validness of the proposed method.
}

The optimal performance achieved by the prediction method in this paper is compared with the predictability upper bound, as shown in the Table. \ref{acc}.
First, the prediction accuracy of the prediction method in the paper reaches the theoretical upper bound of predictability. Then, the measure of predictability can guide the prediction and foresee the accuracy of the prediction that the data set can achieve. Finally, by setting the quantization interval $T$, it is possible to adjust between the tolerable error range and the prediction accuracy.

\linespread{1.3}
\begin{table}
\centering
\caption{The prediction accuracy and predictability of users' voice traffic.}
\label{acc}

\vspace{0.5mm}

\begin{tabular}{cccc}
\Xhline{1pt}
\specialrule{0em}{-1pt}{3pt}
Quantization interval(T) & Model & Prediction accuracy & Predictability \\
\midrule
120s & MC(25) & 71.45\% & 0.70 \\
300s & MC(25) & 76.52\% & 0.75\\
600s & MC(25) & 83.47\% & 0.81\\
\Xhline{1pt}
\end{tabular}
\end{table}

\section{Conclusions}
\label{sec6}

In this paper, several important insights are carried out.
First is the validity of predictability measures for traffic forecasting.
In addition to human mobility prediction, our work demonstrates that traffic prediction can also be measured by entropy and predictability.
Because entropy mines the pattern of subsequences of arbitrary lengths of time series, the scope of the entropy includes both the short term and long term correlation prediction problems. This insight is a significant extension to the original theory, which brings more possibilities for future applications.
Secondly, the traffic quantification methods is proposed and discussed. Based on the real world voice traffic data, the prediction of N-order Markov models, diffusion based model and MF model are presented, among which, 25-order Markov models performs best. Under the coarse-grained quantization of T=600s, the average accuracy of the user set is up to 83.47\%. This exhibits that, the predictability can be achieved by classical models in voice traffic prediction.

Last but not least, by setting the quantization interval $T$ and observing the predictability of quantification, the tolerable error range and expected prediction accuracy can be easily adjusted. Through this, it is expected to find a compromise between the complexity of the model and the accuracy of describing the flow characteristics to solve the difficult problem of traffic prediction.


\section*{Acknowledgment}
This work was partially supported by Key Program of Natural Science Foundation of China under Grant(61631018), Huawei Technology Innovative Research.

\balance

\end{document}